\def\ANON{0}
\def\IEEE{0}
\def\ACM{0}
\def\ARXIV{1}
\def\LNCS{1}
\def\SCITEPRESS{0}

\ifnum\IEEE=1
\documentclass[compsoc, conference, a4paper, 11pt, times]{IEEEtran}
\fi
\ifnum\ACM=1
\documentclass[sigconf]{acmart}
\fi
\ifnum\LNCS=1
\documentclass[a4,11pt,english,runningheads]{llncs}
\fi
\ifnum\SCITEPRESS=1
\documentclass[a4paper,twoside]{article}
\fi

\usepackage{amsmath}
\usepackage{amssymb}
\usepackage{hyperref}
\usepackage{xcolor}

\definecolor{codegreen}{rgb}{0,0.6,0}
\definecolor{codegray}{rgb}{0.5,0.5,0.5}
\definecolor{codepurple}{rgb}{0.58,0,0.82}

\usepackage{url}
\newcommand{\ignore}[1]{{}}

\ifnum\IEEE=1
\newtheorem{lemma}{Lemma}
\fi
\ifnum\SCITEPRESS=1
\newtheorem{lemma}{Lemma}
\fi

\ifnum\ACM=1

\AtBeginDocument{%
  }

\copyrightyear{2023}
\acmYear{2023}
\setcopyright{acmlicensed}
\acmConference[ARES 2023]{The 18th International Conference on Availability, Reliability and Security}{August 29-September 1, 2023}{Benevento, Italy}
\acmBooktitle{The 18th International Conference on Availability, Reliability and Security (ARES 2023), August 29-September 1, 2023, Benevento, Italy}
\acmPrice{15.00}
\acmDOI{10.1145/3600160.3605040}
\acmISBN{979-8-4007-0772-8/23/08}

\fi

\ifnum\SCITEPRESS=1
\usepackage{SCITEPRESS}
\fi

\begin{document}


\title{Cryptanalysis of a Privacy-Preserving Ride-Hailing Service from NSS 2022}

\ifnum\ANON=1
\institute{}
\author{}

\else

\ifnum\ACM=1
\author{Srinivas Vivek}
\email{srinivas.vivek@iiitb.ac.in}
\orcid{0000-0002-8426-0859}
\affiliation{
\institution{International Institute of Information Technology Bangalore}
\city{Bengaluru}
\country{India}
}


\fi
\ifnum\ARXIV=1
\author{
Srinivas Vivek
\thanks{ The ORCID ID of S. Vivek is {0000-0002-8426-0859}. S. Vivek is with the International
Institute of Information Technology Bangalore, Bengaluru 560 100, India. Email: srinivas.vivek@iiitb.ac.in.   
}
}
\fi

\fi

\ifnum\ANON=0
\author{Srinivas Vivek}
\institute{IIIT Bangalore, IN\\
\email{srinivas.vivek@iiitb.ac.in}}
\markboth{}%
{S. Vivek: Cryptanalysis of a Privacy-Preserving Ride-Hailing Service}
\fi


\ifnum\SCITEPRESS=1
\fi

\ifnum\ACM=1

\fi

\pagestyle{plain}

\maketitle

\begin{abstract}
Ride-Hailing Services (RHS) match a ride request initiated by a rider with a suitable driver responding 
to the ride request. A Privacy-Preserving RHS (PP-RHS) aims to facilitate ride matching while ensuring the 
privacy of riders' and drivers' location data w.r.t. the Service Provider (SP). At NSS 2022, Xie et al. proposed a PP-RHS. In this work, 
we demonstrate a passive attack on their PP-RHS protocol. Our attack allows the SP to completely recover the locations of the rider as well as that of the responding drivers 
in every ride request. Further, our attack is very efficient as it is independent of the security parameter.
\end{abstract}

%
%

\ifnum\IEEE=1
\fi

\ifnum\LNCS=1
\fi


\section{Introduction}

Ride-Hailing Services (RHS) such as Uber, Ola, Lyft, Didi, etc. have become very popular in 
recent times. While traditional RHSs need access to the location data and other personal information of riders and drivers to determine ride matching and other related services, this set up also entails privacy risks for rider and drivers. Recent years have witnessed many Privacy-Preserving Ride-Hailing Service (PP-RHS)/ Privacy-Preserving Ride-Sharing Service (PP-RSS) proposals where the main aim of such protocols is to offer ride-matching service by a Service Provider (SP) like in a traditional RHS protocol. Additionally, a PP-RHS is also expected to (at least) ensure the privacy of location data of riders and users from the eyes of the SP, and from each other as well (see \cite{Pham2017PrivateRideAP,ORidePaper}, \cite{pRideLuo,lpRideYu,wangTrace
,pRideHuang2021,yuPSRide,XieGJ21}, and references there in). 

Recent works \cite{Vivek23,MurthyV22,MurthyV22nss,Vivek21,KumaraswamyV21,KumaraswamyMV21} 
have also exposed serious privacy leakages in some of the above mentioned PP-RHS/ PP-RSS protocols. For 
example, the works \cite{MurthyV22,KumaraswamyV21} demonstrate how a rider could passively deduce 
location information of the drivers who respond to its ride request and not just of the driver 
eventually selected to offer the ride in the ORide PP-RHS protocol \cite{ORidePaper}. However, the SP 
would continue to not be able to learn any location information as it would be working only with 
homomorphically encrypted data. Needless to say, the above cited attacks differ in technical details 
to account for varied designs of the underlying protocols potentially employing different cryptographic primitives.\\ 

\noindent\textbf{PP-RHS Protocol from \cite{XieCGLJ22}}.
The privacy-preserving ride-hailing service proposed by Xie et al. \cite{XieCGLJ22} 
was designed to offer 
private ride matching. The protocol enabled the SP to privately compute the distance between the rider (a.k.a. passenger) and the responding drivers (a.k.a. taxis) but without the SP actually learning 
the location coordinates of riders and drivers. The location coordinates of the drivers are 
 encoded using the Road Network Embedding (RNE) technique \cite{shahabiRNE}. This technique enables us to encode the topology of the road network into a high dimensional vector while still allowing the calculation of the actual road distance merely by computing the difference between the two RNE-encoded vectors and then finding the maximum amongst them. For every ride request by a rider, the driver with the least distance to the rider is chosen among the responding drivers.\\ 

\noindent\textbf{Threat model in \cite{XieCGLJ22}}. In \cite[Section 4.2]{XieCGLJ22}, the authors describe the adversarial model under which they argue about the security of their protocol.
The SP is assumed to be \textit{honest-but-curious}, 
i.e., the SP performs its computation as part of the protocol correctly but may be eager to learn 
riders' and drivers' location information from the protocol transcript it possesses. Further, it is 
assumed that the riders and drivers behave responsibly in the sense that they keep their secret 
keys secure.\\

\noindent\textbf{Security claim in \cite{XieCGLJ22}}. The 
authors claim  that SP will not be able to learn any information about
the RNE location coordinates of riders and drivers. Their argument is based on the security of two 
PRFs whose output is used to mask the actual RNE coordinates. See \cite[Section 6]{XieCGLJ22} for 
their security analysis.\\  

\noindent\textbf{Our contribution}: 
We demonstrate that the above security claim by the authors of \cite{XieCGLJ22} is 
\textit{incorrect}. 
Precisely, we exhibit a passive attack
on the PP-RHS protocol from \cite{XieCGLJ22} where an honest-but-curious SP can completely recover 
the RNE coordinates (and eventually 
the actual locations) of the rider and all the responding drivers in every ride request.

Note that the authors of \cite{XieCGLJ22} had incorporated specific security features into their design that would thwart an attack similar to the one in \cite{Vivek23} on the protocol of 
\cite{XieGJ21}. This includes limiting the total amount of data received by the SP by limiting the ride matching to a single zone, matching rides in the same time slot and by limiting the number of driver ciphertexts by implicit matching of the ciphertexts. Our main observation is that in spite of the above countermeasures, the decrypted data that the SP currently sees is enough to mount the attack of \cite{Vivek23}.

\section{PP-RHS Protocol from \cite{XieCGLJ22}}
We briefly recall the protocol from \cite{XieCGLJ22}. For brevity, we only recap those details that are
necessary to describe our attack. See \cite[Section 5]{XieCGLJ22} for all the details.

The RNE encoding of a rider/driver's location node $u$ is denoted as 
\[
E(u) = (E_1(u), E_2(u), \ldots, E_n(u)),
\] where 
$n$ is a chosen parameter and each $E_i(u)$ is an unsigned integer. Let $P$ be 
the (passenger) rider requesting a ride and $T_k$ be the $k$\textsuperscript{th} (taxi) driver who responds to this ride 
request. The RNE 
distance between $P$ and $T_k$ is calculated as 
\[
\max_{i=1}^{n} |E_i(u_P) - E_i(u_{T_k})|.
\] 

For efficiency, each coordinate $E_i(u_P)$ of the rider is further 
decomposed into $m$ many blocks $[E_i(u_P)]_j$ (for $j=0,1,2,\ldots,m-1$) of $l$ many 
bits each. The typical values are $l=1$, $l=2$ or $l=4$. Hence, 
\[
E_i(u_P) = \sum_{j=0}^{m-1} [E_i(u_P)]_j \cdot (2^l)^{j}.
\]
Let us denote the $l$-bit block weight $(2^l)^j$ in the above equation as $w_j$.

To facilitate ride matching on encrypted location blocks, instead of directly encrypting the blocks $[E_i(u_P)]_j$, what are actually encrypted are the following weighted differences between $[E_i(u_P)]_j$ and \textit{every} $l$-bit value $q$:
\[
[E_i(u_P)]_{j,q} = \left( q - [E_i(u_P)]_j \right) \cdot w_j, 
\]    
where $q \in \{0,1,\ldots,2^l-1\}$.

Let $[E'_i(u_P)]_{j,q}$ denote the encryption of  $[E_i(u_P)]_{j,q}$. It is computed as
\[
[E'_i(u_P)]_{j,q} = (C_1,C_2),
\]
where
\[
C_1 = \left(\; \text{tag}_{i,j,q}, F(H(\kappa_1,q||i||j||z||s),\gamma_j) \;\right),
\]
and
\[
C_2 = F(H(\kappa_2,q||i||j||z||s),\gamma_j) \; \oplus \; ((q - [E_i(u_P)]_j) \cdot w_j).
\]
In the above two equations, $\text{tag}_{i,j,q}$ is an identifier (which in turn could be computed from other components), $||$ and $\oplus$ are the concatenation and the xor operators, respectively. $H$ and $F$ are PRFs, $\gamma_j$ is a random nonce that is common for all the $2^l$ ciphertexts corresponding to the block $[E_i(u_P)]_j$. The secret keys $\kappa_1$ and $\kappa_2$ are shared among riders and drivers, and it is distributed by an independent and trusted \textit{key manager}. The rider sends these ciphertexts in a \textit{randomly permuted order} to SP for eventual ride matching.

Let $u_{T_k}$ denote the location of the $k$\textsuperscript{th} responding driver ${T_k}$ for the ride request of the rider $P$.
Let the corresponding decomposition of the $i$\textsuperscript{th} coordinate of the RNE encoding $E(u_{T_k})$ be
\[
E_i(u_{T_k}) = \sum_{j=0}^{m-1} [E_i(u_{T_k})]_j \cdot (2^l)^{j}.
\]
Unlike the case of riders where $2^l$ many ciphertexts are generated corresponding to all the weighted differences, in the case of drivers, only one ciphertext is generated. This is done mainly from the security point of view; in order to limit the amount of data seen by the SP to thwart attacks similar to \cite{Vivek23}. Let $[E'_i(u_{T_k})]_j$ denote the encryption of the block $[E_i(u_{T_k})]_j$. It is computed as
\[
[E'_i(u_{T_k})]_j = (C'_1,C'_2),
\]
where
\[
C'_1 = H(\kappa_1,[E_i(u_{T_k})]_j||i||j||z||s),
\]
and
\[
C'_2 = H(\kappa_2,[E_i(u_{T_k})]_j||i||j||z||s).
\]
Note that all the parameters above are as defined previously for the case of the rider $P$. In particular, the zone ID $z$ and the time slot $s$ must be the same as that for the rider. The driver then sends these ciphertexts in a \textit{randomly permuted order} to SP along with the random nonces $\gamma_j$.

To calculate the RNE distance between $E(u_P)$ and $E(u_{T_k})$, the SP works on encrypted $l$-bit blocks at a time. It ``matches'' the ciphertexts $[E'_i(u_P)]_{j,q}$ of the rider for every $q$ with  $[E'_i(u_{T_k})]_j$. This is done by checking the following condition:   
\[
C_1[2] \overset{?}{=} F(C'_1,\gamma_j). 
\]
The above condition corresponds to

\begin{align*}
F(H(\kappa_1,q||i||j||z||s),\gamma_j) \overset{?}{=} & \\
 F(H(\kappa_1,[E_i(u_{T_k})]_j||i||j||z||s),\gamma_j). & 
\end{align*}
It is easy to see that the above condition holds  if $q = [E_i(u_{T_k})]_j$. With very high probability the SP can now compute
\[
([E_i(u_{T_k})]_j - [E_i(u_P)]_j)*w_j
\]
by unmasking $C_2$ with $F(C'_2\,\gamma_j)$, and eventually compute  
\[
E_i(u_{T_k}) - E_i(u_P).
\]
The authors of \cite{XieCGLJ22} claim that the ciphertext matching condition above holds if and only if $q = [E_i(u_{T_k})]_j$.  But this is incorrect as the functions $F$ and $H$ are only assumed to be PRFs but not collision-resistant hash functions.

Finally, the SP chooses the driver with the minimum RNE distance to offer ride to the rider $P$.

As seen above, it is important to note that the SP actually learns  $[E_i(u_{T_k})]_j - [E_i(u_P)]_j$ (as it knows $w_j$) as plaintext.  

\section{Our Attack}

Our main idea is to use the (signed) differences $[E_i(u_{T_k})]_j - [E_i(u_P)]_j$ between the rider's and many drivers' 
individual $l$-bit blocks that the SP learns in every ride request to eventually recover $E(u_{T_k})$ and $E(u_P)$ as was done in \cite{Vivek23} against the protocol of \cite{XieGJ21}.

The authors of \cite{XieGJ21} were aware of the attack in \cite{Vivek23}. Hence, they specifically included features to enhance security like matching ciphertexts restricted to the same zone ID and the time slot, as seen in the previous section. This was done to limit the number of intermediate values that the SP sees during a ride request.

Our main observation is that there will be many drivers in the same zone and operating in the same time slot who will respond to a ride request. This is particularly true for densely populated areas. Further, from the economics point of view, the SP would choose the zone size and other parameters in order to keep demand and supply in fixed proportions. Hence, it is a less likely event that only a few drivers would respond to a ride request. 
Hence, the above observation implies that we can still mount the attack of \cite{Vivek23} on the current protocol. For the sake of completeness, we recollect the details of the attack from \cite{Vivek23} adapted to the current setting. 

For a given block $[E_i(u_P)]_j$, its differences 
\[
[E_i(u_{T_k})]_j - [E_i(u_P)]_j
\] 
with the corresponding 
blocks for all the responding drivers are collected by the SP (i.e., as $k$ varies over all the responding drivers). If the SP can get
$2^l$ many drivers each having a different value for the corresponding block, then we can apply
the below Lemma \ref{lem:zx} from \cite[Section IV.A]{XieGJ21}.
\begin{lemma}
\label{lem:zx}
Let $x \in \{0,1,2,\ldots,2^l-1\}$. Given all the (signed) differences $\{z-x\;|\; 0 \le z \le 2^l-1 \}$ (in any order), then we can uniquely recover $x$. 
\end{lemma}
Basically, the above lemma allows SP to uniquely recover the value $[E_i(u_P)]_j$. Once $[E_i(u_P)]_j$ is recovered, then it can easily compute $[E_i(u_{T_k})]_j$ from the difference $[E_i(u_{T_k})]_j - [E_i(u_P)]_j$ it already knows. Hence, the SP will be able to recover the locations of the rider as well all the responding drivers in every ride request.

Note that our attack is very efficient since it only deals with plaintext values, hence, it is independent of the security parameter. Also, note that the values of the block length is typically $l=1$, $l=2$ or $l=4$ for efficiency reasons. This is because for large values of $l$, the number of ciphertexts to be generated by the rider will also be large, more precisely, it is $2^l$. As estimated in \cite{Vivek23}, the expected number of drivers who need to respond to the ride request to mount this attack is given in the Table \ref{tab:exp}.  

\begin{table}[h]
\centering
\caption{Expected number of drivers needed to successfully mount the attack for different block sizes $l$ \cite{Vivek23}.} 
\label{tab:exp} 
\begin{tabular}{ |c|c| } 
 \hline
 $l$ & Expected no. of drivers\\
 \hline\hline 
 1 & 3\\
 \hline 
 2 & 9\\ 
 \hline
 3 & 22\\ 
 \hline
 4 & 55\\
 \hline
\end{tabular}
\end{table}

\section{Conclusion}
We have demonstrated a passive attack where an honest-but-curious service provider can completely recover the location of rider and drivers in every ride request in the protocol from \cite{XieCGLJ22}. Our attack was possible by combining data from multiple drivers who respond to a ride request. Interestingly, our attack was possible in spite of designers of \cite{XieCGLJ22} incorporating security features that aimed to prevent the attack of our type.

\ifnum\ANON=0

\ifnum\LNCS=1
\section*{Acknowledgments}
This work was partially funded by the Privacy-Preserving Data Processing and Exchange for Sensitive Data in the National Digital Public Infrastructure (P3DX) project, and the Infosys Foundation Career Development Chair Professorship grant for the author. 
\fi

\fi

\ifnum\IEEE=1
\bibliographystyle{IEEEtran}
\fi

\ifnum\ACM=1
\bibliographystyle{ACM-Reference-Format}
\fi

\ifnum\LNCS=1
\bibliographystyle{alpha}
\fi

\ifnum\SCITEPRESS=1
\bibliographystyle{alpha}
\fi

\bibliography{refs.bib}


\end{document}